\documentclass[twocolumn,showpacs,preprintnumbers,amsmath,amssymb]{revtex4}

\usepackage{graphicx}
\usepackage{dcolumn}
\usepackage{bm}

\begin{document}


\title{One more observational consequence of many-worlds quantum theory
}

\author{A.V. Yurov}
\email{artyom_yurov@mail.ru} \affiliation{%
Theoretical Physics Department, I. Kant's State University of
Russia, A. Nevskogo str., 14, 236041,
 Russia.
}
\author{V.A. Yurov}%
 \email{valerian@math.missouri.edu}
 \affiliation{%
Department of Mathematics, University of Missouri-Columbia, MO
65211, USA
}%


\date{\today}

\begin{abstract}
Using new cosmological doomsday argument Page predicts that the
maximal lifetime of de Sitter universe should be $t_{max}=10^{60}$
yr which is way too small in comparison with strings predictions
($t_f>$googleplex). However, since this prediction is dependant on
the total number of human observations, we show that Page
arguments results instead in astounding conclusion that this
number is the quantum variable and is therefore much greater then
Page's estimation. Identifying it with the number of
coarse-grained histories in de Sitter universe we get the lifetime
of the universe comparable with strings predictions. Moreover, it
seems that this result can be considered as another one of the
observational evidences of validity of the many-worlds quantum
theory. Finally, we show that for the universe filled with phantom
energy $t_{max}\sim t_f$ up to very high precision.

\end{abstract}

\pacs{98.80.Cq;98.80.-k}
\maketitle

\section{\label{sec:level1}Introduction}

In article \cite{Page} Don Page has presented the forcible
argument that the lifetime of the de Sitter universe
$t_{max}<10^{60}$ yr. On the other hand, the string theory
prediction grants the dS universe as much time as $t_f<{\rm
recurrence\,\,time}\sim {\rm e}^{0.5\times 10^{123}}$ yr
\cite{Lifetime1}, \cite{Lifetime2}  (the matter of whether it
should be seconds, years or even millenniums is really unessential
for such monstrous numbers). It is possible to lower this value to
the $t_f\sim{\rm e}^{10^{19}}$ yr and even to the limit of
$t_f\sim{\rm e}^{10^{9}}$ yr for models with instantons of Kachru,
Pearson and Verlinde \cite{KPV} and with 2 Klebanov-Strassler (see
\cite{KS}) throats \cite{FLW}. But, nevertheless, even with
assumption that one of those models do describe our Universe, the
magnitude $t_f$ will still be way too large as compared to Page's
$10^{60}$ yr. Thus we are facing the following dilemma: either the
dark energy is not pure positive cosmological constant at all (and
stringscape should not have {\em any} significant long-lived
positive metastable minima) or the dark energy IS the cosmological
constant and our ideas about stringscape (and the strings theory
in general) are absolutely false!

Actually, the nature of dark energy is one of those questions, which can be redirected to astronomers.
It appears, that there exist some observational series which proved to be sufficiently difficult to explain with
the assumption that the dark energy should be some scalar field (quintessence)
rather then cosmological constant. (The phenomenon of this kind is,
for example, the drift of unhomogendous local volume (1 Mpc) with
the regular Hubble flow inside \cite{Chernin}). Of course, those results can appear to be of
statistically insignificant nature,  but if not, then it would mean one strong graphic evidence of presence of the
cosmological constant.

So, does there exist some kind of ''loophole'' in the Page reasoning?
Something, that would allow us to conclude that the lifetime of dS universe can be
$t_{max}\sim t_f$ with $t_f\sim{\rm googleplex}$? As we shall show further, such loophole indeed exists.
To present it we will have to re-examine the essense of the Page's argumentation, and it will be done in
the next Section. In Section III we'll consider the case of
phantom cosmology to show that it surprisingly grants us
$t_{max}=t_f$, and therefore, in such universes the Page reasoning doesn't lead to
inevitable conflict, as differs from dS models. Next Section is essentially devoted to the universes filled with
vacuum energy and to the way of preventing the Page conclusion $t_{max}\ll
t_f$. Here we will show that it is possible to obtain $t_{max}\sim t_f$ in assumption
that the total number of human observations is the quantum variable. And Section V is the overall conclusion.

\section{Page Argument}

Following to Page, suppose that the process of observation is described by some localized
positive operator $A$, such that application of it to any state $\psi$ leads to positive central
tendency. This implies that every possible observation has some
positive probability of occurrance in the given volume (e.g., as a vacuum fluctuation).
Therefore, we can treat the observers as the standart quantum objects.
With this in mind, Page has calculated the action for the brain of a human
observer: $S_{{\rm br}}\sim10^{16}$ $J\times s$, and the
probability $p_{{\rm br}}\sim {\rm e}^{-S/\hbar}\sim {\rm
e}^{-10^{50}}$. Then, Page made an estimation for 4-volume for the brain
($V_4({\rm br})$), taken in process of making the observation: $V_4({\rm
br})\sim{\rm e}^{331}\,a_{_{\rm Pl}}^4$.

Next, let's assume that we are living in dS universe filled with
vacuum, energy density $\rho_{_\Lambda}\sim 10^{-29}\, {\rm
gramme/cm^3}$ of which greatly exceeds the total density of all other energy
components in the universe. Then we appears to be merely prisoners in
"cosmic prison" of a radius $R=c/H$, where Hubble constant
$H=\sqrt{8\pi G \rho_{_\Lambda}/3}$. After $10^{17}$ yr each and every star in the universe
will be either black hole, black dwarf or neutron star; $10^{10}$ Gyr
later the temperature of neutron stars will decrease to less than
$100\,\,K$. It is mildly speaking unlikely that human-observers will be able to endure in such
inhospitable universe. However, it wouldn't really matter at that point, because no life (including human-observers)
will be able to exist there forever due to both proton decay (it's time life $t_{{\rm
pr}}>10^{32}$ yr) and the exponential falloff of the density of
matter (information, being processed in ever-expanding universes was
considered in \cite{Barrow}. There has been shown that an infinite amount
of information can be processed via the usage of temperature
gradients created by gravitational tidal energy, but only in assumption
that the cosmological constant is equally zero). Therefore if
$t_f\sim{\rm googleplex}$ then except for unimaginably tiny initial
period from the big bang to $t_{{\rm pr}}$ the universe will be absolutely
dead. It is definitely not bright future at all!

On the other hand, taking into account the unimaginably long lifetime
of such universe we shall conclude that all possible events, including
those with extremely low probability, will someday occur. One of the
most interesting of those unlikely events would be the spontaneous
appearance from quantum fluctuations of ''observers'', surrounded by
''environment'' suitable to permit the ''observation''.
With this conclusion, it would be only natural to ask: can we in principle be one of those ''vacuum
observers''? And, more generally: under what circumstances will the ordered (i.e. classical) observations
dominate over vacuum ones? Page gives the following answer: if
$t<t_{max}=10^{60}$ yr, and only then will the human observations be with high probaility
ordered. Otherwise, almost all observations in the universe will have
its root in vacuum fluctuations. As the result, in universe with $t_{max}\sim t_f\sim {\rm
googleplex}$ our, obviously ordered, observations are to be considered as something embarassingly
atypical. Page concludes that '' This extreme atypicality is
like an extremely low likelihood, counting as very strong
observational evidence against any theory predicting such a
long-lived universe with a quantum state that can allow localized
observations'', and makes the prediction that the universe just will not last
long enough to give 4-volume $>{\rm e}^{10^{50}}$.

To show this in work, let's consider the total 4-volume of universe:
\begin{equation}
V_4(t)=c\int_{0}^{t} dt a^3(t). \label{V4}
\end{equation}
The probability of vacuum fluctuations $p_{{\rm vac}}<p_{{\rm
br}}$ whereas the probability of ordered occurrences $p_{{\rm
ord}}>p_{{\rm br}}$. Multiplying $V_4(t)$ by  $p_{{\rm ord}}$ results in
the volume of the part of total $V_4(t)$ where ordered occurrences are dominating ones.
Now let $N$ be the number of observations, made during the past human
history. The product $N V_4({\rm br})$ will mark the part
of total $V_4(t)$ where ordered human observations all take place. If
humans are the typical observers (anthropic principle!) then one can expect that
\begin{equation}
V_4(t) p_{{\rm ord}}\sim V_4({\rm br})N. \label{urav}
\end{equation}
Substituting  $a(t)=a_0{\rm e}^{Ht}$ into the (\ref{V4}) one get
$V_4(t)$. Following Page we can evaluate $N\sim{\rm e}^{48}$. Substituting $N$
and $V_4(t)$ into the (\ref{urav}) allows us to express $ p_{{\rm
ord}}$. Finally, using the inequality $p_{{\rm ord}}>p_{{\rm br}}$
one comes to conclusion that, under those circumstances, the timelife of the dS
universe is indeed $t<t_{max}=10^{60}$ yr.

\section{Phantom energy}

Let's see, what will happen with Page results in the universe filled with phantom
energy. It appears, that in contrast to dS models, for such universes
we get a remarkable concordance: $t_f=t_{max}$ up to very high
degree of accuracy.

Before we start, we should mention, that the particular interest to the models
with phantom fields arises from their prediction of so-called "Cosmic Doomsday" alias
big rip \cite{1}. Since for the phantom energy we have
$w=p/(c^2\rho)=-1-\epsilon$ with $\epsilon>0$, the integration of the
Einstein-Friedmann equations for the flat universe results in
\begin{equation}
\begin{array}{cc}
\displaystyle{
a(t)=\frac{a_0}{\left(1-\xi t\right)^{2/3\epsilon}},}\\
\displaystyle{
\rho(t)=\rho_0\left(\frac{a(t)}{a_0}\right)^{3\epsilon}=\frac{\rho_0}{(1-\xi
t)^2}}, \label{1}
\end{array}
\end{equation}
where $\xi=\epsilon\sqrt{6\pi G\rho_0}$. Choosing $t=0$ as the
present time, $a_0\sim 10^{28}$ cm and
$\rho_0=1.4\rho_c/(2+3\epsilon)$ as the present values of the
scale factor and the density (If $\epsilon\ll 1$ then $\rho_0\sim
0.7\times 10^{-29}\,{\rm g}/{\rm cm}^3$), at time $t=t_f=1/\xi$,
we automatically get the big rip.

Now, let's return to our question. Equations (\ref{V4}) and (\ref{1}), taken together, lead to
$$
\displaystyle{
V_4(t)=\frac{ca_0^3\epsilon}{\xi(2-\epsilon)}\left(\frac{1}{\left(1-t/t_f\right)^{(2-\epsilon)/\epsilon}}-1\right)+V_4(0),}
$$
where $V_4(0)=a_0^4=10^{112}\,{\rm cm}^4={\rm e}^{258}\,{\rm
cm}^4$. Using Page approach we have
$$
\displaystyle{
\frac{ca_0^3\epsilon}{\xi(2-\epsilon)}\left(\frac{1}{\left(1-t/t_f\right)^{(2-\epsilon)/\epsilon}}-1\right)<
V_4({\rm br}){\rm e}^{S_{\rm br}/\hbar}-V_4(0).}
$$
The second member of the equation is
$$
3\times {\rm e}^{48}\times 10^{12}\times{\rm e}^{10^{50}}-{\rm
e}^{258}\sim {\rm e}^{10^{50}},
$$
therefore
\begin{equation}
\left(1-\frac{t}{t_f}\right)^{(\epsilon-2)/\epsilon}<{\rm
e}^{10^{50}}. \label{neravenstvo}
\end{equation}
In the case $\epsilon\ll 1$ we get
\begin{equation}
\displaystyle{ \frac{t_f-t}{t_f}>{\rm exp}\left(-\frac{0.5\times
10^{50}}{t_f\sqrt{6\pi G\rho_0}}\right)={\rm
exp}\left(-\frac{1.685\times10^{67}}{t_f}\right)}.
\label{neraven-2}
\end{equation}
Now, we have to consider 3 different cases.
\newline \newline
{\em a}. $t_f\gg 1.685\times 10^{67}\,{\rm s}=5.3\times
10^{59}\,{\rm yr}$. In this case the power of exponent in
(\ref{neraven-2}) is small enough to use the expansion in
Taylor's series. It's application results in inequality
$t<t_{max}=5.3\times10^{59}\,{\rm yr}$. This leaves us with the same
problem as in dS situation: the end of the world will take place at
$t=t_f$ but ordered observation will be dominating ones while $t\ll
t_f$ only.
\newline \newline
{\em b}. $t_f\ll 1.685\times 10^{67}\,{\rm s}$. In this case
$$
\displaystyle{ t<t_{max}=t_f\left(1-{\rm e}^{-1.685\times
10^{67}/t_f}\right)\sim t_f.}
$$
Here we come to remarkable difference between phantom and dS
cosmologies. While in the last case we have $t_{max}=10^{60}\, {\rm yr}\ll
t_f>{\rm googleplex}$, where $t_{max}$ follows from Page's reasoning
and $t_f$ is the string theory prediction, in the former case the situation can be much more
agreeable: in fact, the validity of the {\em b} condition ensures that $t_{max}\sim t_f$.
\begin{table}
\caption{\label{tab:table2}
}
\begin{ruledtabular}
\begin{tabular}{cc}
$t_f\,{\rm yr}$& $t_{max}/t_f$ \\
\hline\hline $2.3\times 10^{59}$& $0.9$\\
$1.1\times 10^{59}$& $0.99$\\
$7.7\times 10^{58}$& $0.999$\\
$5.8\times 10^{58}$& $0.9999$\\
$4.6\times 10^{58}$& $0.99999$
\end{tabular}
\end{ruledtabular}
\end{table}
As we can see from Tabl. I, $t_{max}\to t_f$ very fast when $t_f$
decreases. If $t_f=5.3\times 10^{50}$ yr then $t_{max}=t_f(1-{\rm
e}^{-10^9})$ and $t_f=22$ Gyr stands for $t_{max}=t_f(1-{\rm
e}^{-10^{50}})$, thus actually erasing the very difference between $t_{max}$ and $t_f$.
\newline \newline
{\em c}. $t_f\sim 1.685\times 10^{67}\,{\rm s}$. This case implies
$$
t_{max}=\left(1-\frac{1}{\rm e}\right)t_f\sim 0.63 t_f.
$$
Therefore, in such Universe only about half of all observers can assuredly consider themselves
classical and having the naturally ordered observations, which is sufficiently better then what we had in dS universe,
yet still being far from perfect.

Summarizing all of the above, we can conclude that the one and essentially the only convenient case is {\em b}. After all, for
$t_f<10^{59}$ yr it gives us $t_f\sim t_{max}$ for granted!

\section{The number of coarse-grained histories}

For the time being, the physical meaning of phantom fields is as
yet unclear. For this reason let's  return back to the realistic
case of dS universe and seeming inconsistency between $t_{max}$
and $t_f$ ($t_{max}\ll t_f$), that has been found in it. The core
of Page's argumentation is the equation (\ref{urav}). But let's
inspect carefully the quantities, forming it. It is clear that, by
complete analogy with $p_{{\rm br}}$, quantity $p_{{\rm ord}}$
should be calculated by quantum laws. Indeed, in the  framework of
Page approach one need make a comparison $p_{{\rm br}}$ with
$p_{{\rm ord}}$. It is as well to remember that $p_{{\rm br}}$ is
the quantum quantity, therefore, generally speaking, the same must
be true for the $p_{{\rm ord}}$. As a matter of fact, $p_{{\rm
ord}}={\rm e}^{-S_{\rm ord}/\hbar}$. Therefore, l.h.s of equation
(\ref{urav}) is dependant on $\hbar$. But the equivalence will
hold only if the same will be true for the r.h.s.! If the value
$V_4({\rm br})$ is purely classical, then $N$ is the only
remaining candidate for the dependancy on $\hbar$.

At a first glance this conclusion seems absolutely grotesque, but it appears to be right in touch with Page
resonings. As a matter of fact, in his article Page deals with quantum (or
semi-classical) observers. The number of quantum observers $N$ is the
quantum quantity and hence, must be calculated by the quantum laws. From this point of view, it is
no wonder that $N$ will depend on $\hbar$.

But if this is correct, then one can't use Page estimate ($N\sim{\rm e}^{48}$) anymore. Unfortunately, we
can't calculate $N$ explicitly, but we can evaluate it upon
usage of very simple quantum-based reasoning. It is already clear that ''new'' $N$ should be much greater then ${\rm e}^{48}$.
As we shall see, this number can exceed even gogleplex, thus totally refuting Page argument.

One can roughly  evaluate the number $N$ as the number of
coarse-grained histories: $N=N_h=N_b^{N_c}$ where $N_c$  is the
number of spacetime cells and $N_b$  is the number of relevant
bins in field space. In the article \cite{Vilenkin} Garriga and
Vilenkin have made this for the spacetime volume with the size $R=ct_0$
where $t_0=10^{10}$ yr. As a result they got $N\sim{\rm
e}^{10^{244}}$. Substituting this value into the (\ref{urav}) one
get $t_{max}=10^{261}$ yr. This number is by many orders greater than
Page's  $10^{60}$ yr but is still too small in comparison with
${\rm e}^{10^{131}}$. However, the number $N$ easily allows for additional incease up to
to the point, where $t_{max}$ will be comparable with
strings predictions.

Indeed, remaining in framework of quantum theory we should consider all possible observers,
including those who are living in much older universes where
vacuum energy already exceeds the total density of all the
other energy components in the universe. In such universes
$$
V_4(t)\sim \frac{ca_0^3}{\sqrt{24\pi G\rho_{_\Lambda}}}{\rm
e}^{\sqrt{24\pi G\rho_{_\Lambda}}}={\rm e}^{0.5\times 10^{-17}t}.
$$
For example, if $t=10^{17}$ yr (the era of black holes) one have
$V_4={\rm e}^{0.2\times 10^8}$  and if $t=10^{32}$ yr (the low
bound of proton lifetime)  then $V_4={\rm e}^{0.2\times 10^{23}}$.
The number of spacetime cells of size $L$  will be $N_c\sim
V_4(t)/L^4\sim {\rm e}^{10^8}$ in first case and $N_c\sim {\rm
e}^{10^{23}}$ in the second one. But in all cases the values of
$N$ are given by ''supergoogleplex'' numbers:
$$
N={\rm e}^{{\rm e}^{10^8}},\qquad N={\rm e}^{{\rm e}^{10^{23}}}.
$$
Substituting them in (\ref{urav}) we finally get $t_{max}={\rm e}^{10^8}$ yr
or $t_{max}={\rm e}^{10^{23}}$ yr.

The interesting fact here is that both these numbers lies in remarkable agreement with the results
of \cite{FLW}. In particular, for the case of 3
${\overline{D3}}$-branes with some parameters there have been obtained theoretical value $t_f\sim {\rm e}^{10^{19}}$
(lifetime on the NS5-brane). Decays due to
decompactification are much faster: $t_f\sim {\rm e}^{10^{17}}$.
Those are results of the models with the single KS throat. Consideration of 2 KS
throats (such models are more interesting since they result in
positive cosmological constant whereas the models with single KS
throat result in $\Lambda<0$) in case of KPV instantons
leads to such value as $t_f\sim {\rm e}^{10^{9}}$ - very good agreement
with the previously obtained $t_{max}={\rm e}^{10^8}$.

In the case of general position one can conclude that
$$
t_{max}=10^{17}{\rm e}^{10^{-17}{\tilde t}}\,\,{\rm yr},
$$
where ${\tilde t}$ is the maximal possible lifetime of
''human-observers''. Thus, if $t_{max}=$goggleplex one get
${\tilde t}=10^{117}$ yr while $t_{max}={\rm e}^{10^{123}}$ yr
implies ${\tilde t}=10^{140}$ yr. Of course, it can be difficult to
imagine that $10^{117}$ yr later the universe will be filled
by ''human-observers''. Besides, it can be argued whether such
''observers'' fits into the set being reviewed or not. But the answer is very
simple: whenever the probability of finding ourselves in such universe has
the nonzero value, we have to take it into account.

Finally, we should answer the following question: are those ''auxiliary'' observers real,
or not, i.e. can we ascribe all of them to some really existing Universes, or
are they nothing more then ''vacuum probabilities''? The answer is: yes, they have to be real; otherwise,
we are facing the situation, where the $10^{10^{100}}$ quantum objects are required to
explain the existence of ${\rm e}^{48}$ (real) objects. Here is the same Page's
paradox, only in other form and aggravated by much worsen numbers!
\section{Conclusions}
As we have seen, the assumption that $N$ is the total number of quantum observers
results in such lifetime of universe which is comparable with
strings predictions. This creates the very strong grounds for serious
consideration of such strange possibility. After all, the quantum
nature of $N$ seems to be absolutely inevitable in quantum cosmology.

Of course, such state of affairs is something highly unusual in ''everyday'' quantum mechanics. It has already become a common fact,
that in laboratory research with neutron interferometer the
neutron passing through a beam splitter will split into ''two
neutrons''. But in lab we don't expect that the same will be true
for us. Observers are classical objects ''ad definition''.

However, in quantum cosmology this situation changes drastically. Since we
are nothing but the part of the universe we have no choice but to consider
ourselves as quantum objects. Page has shown in \cite{Page1} that
quantum cosmology can give observational consequences of
many-worlds quantum theory. We think that our results can be
consider as one more observational  evidence of validity of
many-worlds quantum theory.

$$
{}
$$

\centerline{\bf REFERENCES} \noindent \begin{enumerate}

\bibliography{apssamp}
\bibitem{Page} Don N. Page, The Lifetime of the Universe,
hep-th/0510003.

\bibitem{Lifetime1} N. Goheer, M. Kleban, and L. Susskind, The Trouble With
De Sitter Space. J. High Energy Phys. 07, 056 (2003),
hep-th/0212209
\bibitem{Lifetime2} S. Kachru, R. Kallosh, A. Linde, S. P. Trivedi, De Sitter Vacua in String
Theory. Phys.Rev. D68 (2003) 046005,  hep-th/0301240.
\bibitem{KPV} S. Kachru, J. Pearson, and H. Verlinde,
Brane/Flux Annihilation And the String Dual Of a
Non-Supersymmetric Field Theory. JHEP 06 (2002) 021,
hep-th/0112197.

\bibitem{KS} I. R. Klebanov and M. J. Strassler, Supergravity And a Confining
Gauge Theory: Duality Cascades and chiSB-resolution of Naked
Singularities. JHEP 08 (2000) 052, hep-th/0007191.

\bibitem{FLW} A. R. Frey, M. Lippert, and B. Williams,
The Fall of Stringy De Sitter. Phys. Rev. D68, 046008 (2003),
hep-th/0305018.
\bibitem{Chernin} A. D. Chernin, Cosmic Vacuum. PHYS-USP. 44 (11),
2001, 1099-1118.
\bibitem{Barrow} J.D. Barrow and  S. Hervik, Information Processing in Ever-expanding
Universes. Phys.Lett. B566 (2003) 1-7, gr-qc/0302076.
\bibitem{1} R.R. Caldwell, M. Kamionkowski and N.N. Weinberg, Phantom Energy and the Cosmic Doomsday.
Phys. Rev. Lett. 91 (2003) 071301, astro-ph/0302506.
\bibitem{Vilenkin} J. Garriga and A. Vilenkin, Many Worlds in One.
Phys.Rev. D64 (2001) 043511, gr-qc/0102010.
\bibitem{Page1} Don N. Page, Observational Consequences of Many-Worlds Quantum Theory,
quant-ph/9904004.








\end{enumerate}

\vfill \eject

\end{document}